\documentclass[aps,prb,twocolumn,superscriptaddress,showpacs]{revtex4}
\usepackage{mathptmx}
\usepackage{graphicx}
\begin{document}
\title{Comment on ``Band structure engineering of graphene by strain:
First-principles calculations''}
\author{M. Farjam}
\affiliation{Department of Nano-Science, Institute for Research in
Fundamental Sciences (IPM), P.O. Box 19395-5531, Tehran, Iran}
\author{H. Rafii-Tabar}
\affiliation{Department of Nano-Science, Institute for Research in
Fundamental Sciences (IPM), P.O. Box 19395-5531, Tehran, Iran}
\affiliation{Department of Medical Physics and Biomedical Engineering
and Research Centre for Medical Nanotechnology and Tissue Engineering,
Shahid Beheshti University of Medical Sciences, Evin, Tehran 19839,
Iran}
\date{\today}

\begin{abstract}
In their first-principles calculations of the electronic band structure of
graphene under uniaxial strain, Gui, Li, and Zhong [Phys.\ Rev.\ B \textbf{78},
075435 (2008)] have found opening of band gaps at the Fermi level. This finding
is in conflict with the tight-binding description of graphene which is closed
gap for small strains.  In this Comment, we present first-principles
calculations which refute the claim that strain opens band gaps in graphene.
\end{abstract}

\maketitle

Gui \textit{et al.} \cite{gui2008} have used first-principles calculations to
investigate the effect of planar strain on the electronic band structure of
graphene, and have found opening of band gaps at the Fermi level resulting from
arbitrarily small uniaxial strains, applied parallel or perpendicular to the C-C
bonds.  However, tight-binding (TB) model on the honeycomb lattice with
different nearest-neighbor hoppings in the three directions has been rigorously
shown to be closed gap as long as the hoppings satisfy the triangle inequality.
\cite{hasegawa2006,wunsch2008} The closed-gap TB model, which contains zero
modes, has been further developed recently to include the effects of magnetic
fields \cite{goerbig2008,dietl2008} and corrugations in graphene.
\cite{wehling2008} The discrepancy between the first-principles calculations of
Ref.~1 and TB model has already been discussed \cite{pereira2009} but the reason
has not been identified. One suggested explanation \cite{pereira2009} is that
this band gap opening is an artifact of density-functional theory (DFT)
calculations.  However, the possibility that the nearest-neighbor TB model is an
incomplete description has not been ruled out.  \cite{pereira2009,kishigi2008}

We remind that the DFT methods essentially solve single-particle Schr\"odinger
equations (Kohn-Sham equations) for effective potentials based on the underlying
lattice, and the TB model solves the same problem in a simplified approximation.
Therefore, it seems unlikely that qualitative differences exist between DFT and
TB band structures.  We also see indications of possible error in Ref.~1.
First, Figs.~3(c) and 5(c) of Ref.~1 show a peculiar peak whose underlying cause
is not explained.  Second, an energy gap is incorrectly ascribed to the TB band
structure which is then plotted in Fig.~4 of Ref.~1 with large symbols that hide
the important band crossing.

In this Comment, we check directly the first-principles calculations of Ref.~1
by one of the available DFT codes.  We used the {\sc Quantum-ESPRESSO}
\cite{espresso} package based on the pseudopotential plane-wave method.  We
obtained the pseudopotential C.pw91-van\_ak.UPF also from Ref.~9 and used a
kinetic-energy cutoff of 40~Ry, a Monkhorst-Pack $k$-point mesh of
$21\times21\times1$, and a vacuum separation of $20.5$~\AA\ along the $c$ axis.
We chose these parameters as close as possible to those of Ref.~1 for a more
meaningful comparison. \cite{vasp}

\begin{figure}[b]
\includegraphics[width=60mm]{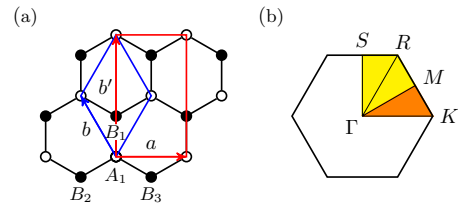}
\caption{\label{fig1} (a) (Color online) Graphene honeycomb lattice.  It is
formed from a triangular Bravais lattice with a two-atom basis consisting of
$A_1$ and $B_1$.  Under uniaxial strain the Bravais lattice becomes centered
rectangular.  (b) First Brillouin zone.  $\Gamma{K}M$ and $\Gamma{K}RS$ enclose
the irreducible wedges corresponding to the triangular and centered rectangular
lattices, respectively.}
\end{figure}

First, we determined the equilibrium lattice constant of graphene in the absence
of strain.  We found a value of $a=2.464$~\AA, defined in Fig.~\ref{fig1}(a),
which is not significantly different from the 2.4669~\AA\ found in Ref.~1.  We
then made calculations on graphene under uniaxial strain for two special cases,
for which Ref.~1 has found maximum values for band-gap openings.  These two
cases are (i) 12.2\% strain applied parallel to the C-C bond, i.e, along $b'$
direction in Fig.~\ref{fig1}(a), and (ii) 7.3\% strain applied perpendicular to
the C-C bond.  For comparison, recent experimental studies of strain in graphene
have applied strains of up to 1.3\% by stretching or bending a flexible
substrate, on which graphene was deposited, and have measured them by Raman
spectroscopy.  \cite{ni2008,ni2009,mohiuddin2009}  Relatively large strains used
in our calculations are more convenient to demonstrate the effects but the
conclusions apply equally to smaller strains.  The uniaxial strain deforms the
triangular lattice of graphene into centered rectangular lattice, shown in
Fig.~\ref{fig1}.  Thus for case (i), $b'$ is fixed at 12.2\% larger value than
its unstrained value of $a\sqrt{3}$, and the value of $a$ is varied until the
stress in the $a$ direction becomes vanishingly small. Of course, for each
choice of $a$, the positions of the atoms must be relaxed until interatomic
forces become sufficiently small.  We found Poisson's ratio to be $\simeq0.10$
for case (i).  A similar procedure is used for case (ii), with fixing $a$ at a
value of 7.3\% larger than the original value, and then optimizing $b'$.  Here
Poisson's ratio was found $\simeq0.14$.

\begin{figure}
\includegraphics[width=85mm]{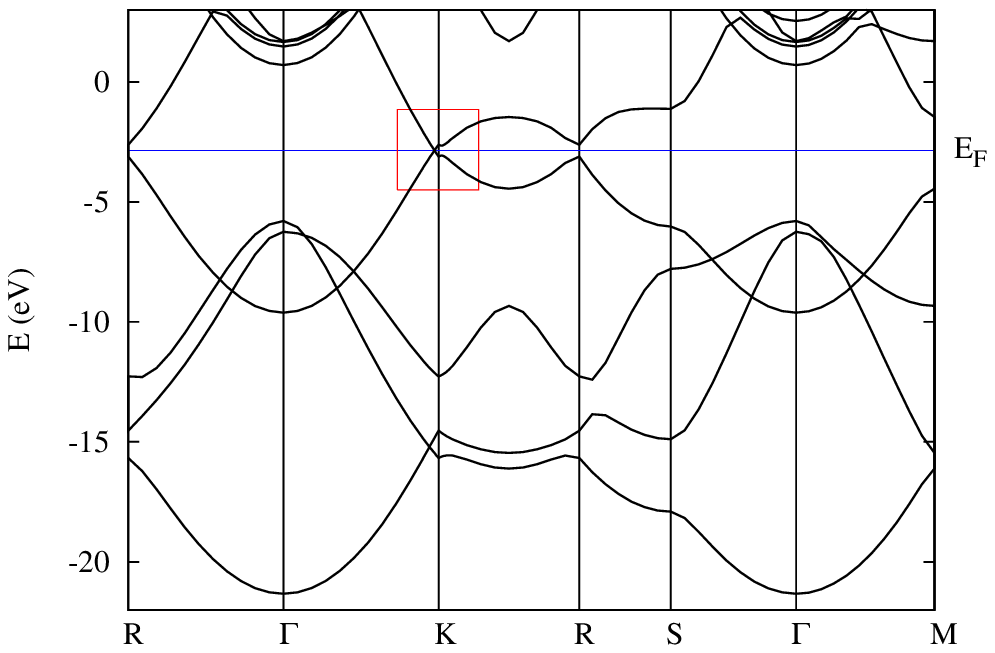}
\includegraphics[width=85mm]{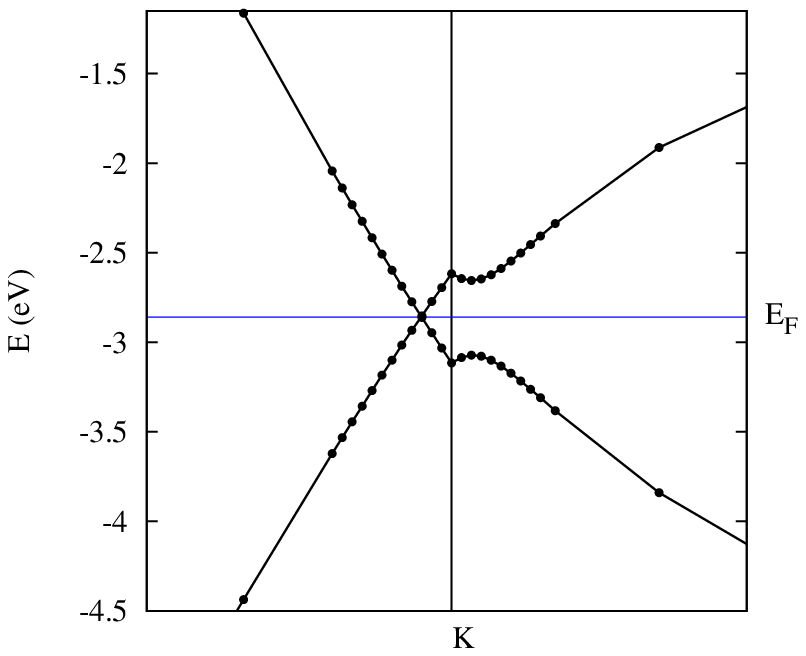}
\caption{\label{fig2} (Color online) Band structure of graphene under 12.2\%
strain applied parallel to C-C bond (top). Magnified portion of band structure
near the contact point (bottom).}
\end{figure}

The band structure we obtained for case (i) along the $k$-point path of
Fig.~\ref{fig1}(b) is shown in the top panel of Fig.~\ref{fig2}, with its
important portion magnified in the bottom panel. In the band plots, we used a
regular $k$-point mesh of 60 points for the entire path and refined it by the
addition of 20 extra points as shown in the magnified part.  The existence of a
contact is clearly seen between the conduction and valence bands near $K$ on
the $\Gamma{K}$ line of the Brillouin zone. The displacement of the Dirac cone
along $\Gamma{K}$ toward $\Gamma$ is in agreement with the TB description of
Ref.~2.  Our Fig.~2 is to be compared with Fig.~3 of Ref.~1, where they give a
value of 0.486~eV for the band gap.  In our calculation, we find a value of
0.498~eV for the energy splittings at $K$ and $R$ in the top panel of
Fig.~\ref{fig2}, which is probably what is taken, in this case, as the band gap
by Ref.~1, having missed the nearby band crossing.

\begin{figure}
\includegraphics[width=85mm]{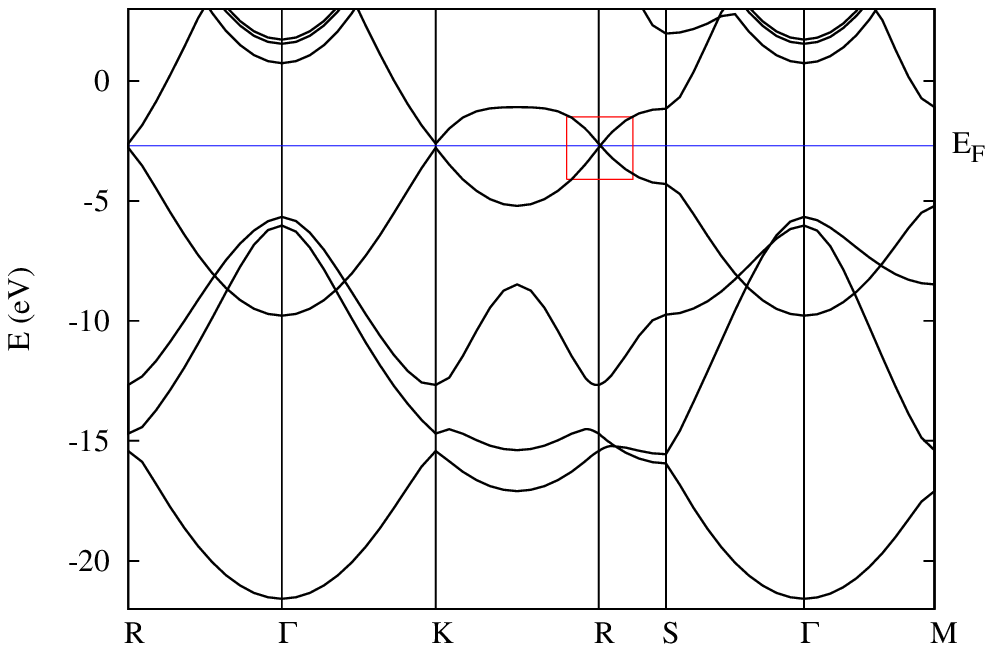}
\includegraphics[width=85mm]{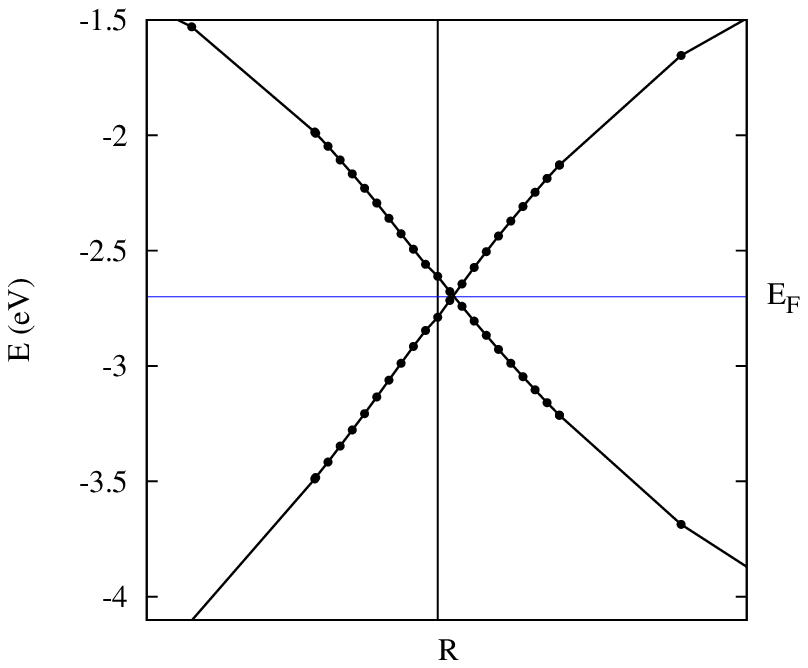}
\caption{\label{fig3} (Color online) Band structure of graphene under 7.3\%
strain applied perpendicular to C-C bond (top). Magnified portion of band
structure near the contact point (bottom).}
\end{figure}

The band structure corresponding to case (ii) is shown in Fig.~\ref{fig3}, and
must be compared with Fig.~5 of Ref.~1.  The main difference with case (i) is
that here the band crossing occurs near the $R$ point on the $RS$ line.  This is
equivalent to a shift of the Dirac point along the $\Gamma{K}$ line away from
$\Gamma$, i.e., in the opposite direction to that of case (i).  In Ref.~1 a
value of 0.170~eV is given for the band gap for this case. We found a value of
0.178~eV for the energy splittings at $R$ and $K$, which is, as in the other
case, close to the band gap given in Ref.~1.

In conclusion, our first-principles calculations establish that graphene under
uniaxial strain is gapless in agreement with the tight-binding model.  Our
numerical values and the general shape of band structures are quite similar to
those found in Ref.~1. However, accidental degeneracies in the band structure
have been disregarded in Ref.~1, and this has resulted in the appearance of
spurious maxima in band gaps as a function of strain.

\section*{ACKNOWLEDGMENTS}
M.F. acknowledges funding from the Iranian Nanotechnology Inititiative,
and H.R.-T. from the Iran National Science Foundation.


\begin{thebibliography}{12}
\bibitem{vasp}
We note that the {\sc vasp} code was used in Ref.~1. However, with correct usage
and interpretation, we expect any standard DFT code to produce the same gapless
spectrum that we obtained.
\end{thebibliography}

\end{document}